# Nanopatterning by Laser Interference Lithography: Applications to Optical Devices

Jung-Hun Seo[2], Jung Ho Park[1], Zhenqiang Ma[2*], Jinnil Choi[3*] and Byeong-Kwon Ju[1*]

[1]Display and Nanosystem Laboratory, School of Engineering, Korea University, Seoul 136-713, Republic of Korea
[2] Department of Electrical and Computer Engineering, University of Wisconsin-Madison, Madison, Wisconsin 53706, USA
[3] Department of Mechanical Engineering, Hanbat National University, Daejeon 305-719, Republic of Korea

A systematic review, covering fabrication of nanoscale patterns by laser interference lithography (LIL) and their applications for optical devices are provided. LIL is a patterning method with simple, quick process over a large area without using a mask. LIL is a powerful technique for the definition of large-area, nanometer-scale, periodically patterned structures. Patterns are recorded in a light-sensitive medium that responds nonlinearly to the intensity distribution associated with the interference of two or more coherent beams of light. The photoresist patterns produced with LIL are the platform for further fabrication of nanostructures and growth of functional materials which are the building blocks for devices. Demonstration of optical and photonic devices by LIL is reviewed such as directed nano photonics and surface plasmon resonance (SPR) or large area membrane reflectors and anti-reflectors. Perspective on future directions for LIL and emerging applications in other fields are presented.

## CONTENTS

## 1. INTRODUCTION

Laser interference lithography (LIL), the maskless exposure of a photoresist layer with two or more coherent light beams, provides a facile, inexpensive, large-area nanolithography technique. Some groups refer to interferometric lithography (IL) interchangeably as interference lithography and holographic lithography [1-7]. The capabilities and techniques of IL have been reviewed elsewhere [8-10] and are only briefly covered here. The purpose of this review is to examine various fields in which IL is impacting nanostructure research. IL shares this large-area capability with advanced optical lithography, today's integrated circuits are at the 22 nm node [11]; nanoimprint lithography [12, 13]; and various self-assembly approaches such as nanosphere lithography and surface plasmon lithography [14-17]. However, each of these patterning techniques has certain limitations and therefore faces difficulties to become a dominant patterning technique. The modern lithography tools for generating nanopatterns are so delicate and expensive that require either highly skilled operators or need to spend lots of budget for maintain the system [18, 19]. Nanosphere lithography [21-24], which uses the spaces between monodisperse colloidal spheres deposited or spun on a densely patterned surface, often lacks long range order, due to the tendency to nucleate uncorrelated domains on the substrate, which yields relatively small (1~100 um) grain boundaries.

Intensive nanoscience research has progressed with serial writing tools such as electron-beam lithography (EBL) [25], ion-beam lithography (IBL) [26-28], and atomic force microscopy (AFM) [29, 30]. Although significant progress has been made, credited to their advanced capability and applicability, investigations on large area remained to be problematic due to their throughput restrictions. A classical throughput value for IBL is that writing 1 Tb of features takes roughly 1 month of continuous exposure time. In comparison, the physics involved with IL process are sufficiently straight forward. The principle is based on the interference between two beams, split from a coherent laser source, forms a standing wave which is recorded on a photoresist coated wafer at angles of $+\theta$ and $-\theta$. The resulting interference pattern has a period of $\lambda/2\sin\theta$ [1]. For the light sources of LIL, lasers with wavelength that closely matches the photoresist, utilized for the semiconductor industries, have been the focus of the research. These wavelengths include 364 and 355 nm, which both match the I-line resists designed for 365 nm; doubled Ar at 244 nm, which matches KrF resists (248 nm); and 213 nm (fifth-harmonic of a Nd:YAG laser) and 193 nm (ArF laser), which match the industry-standard deep UV resists (developed for 193 nm). The minimal feature sizes possible are 182, 178, 122, and 96 nm, respectively. Further reductions in scale are possible using immersion techniques. Most of the effort on immersion IL to date has been at the ArF wavelength to generate minimal pattern sizes.

LIL is very innovative in nanolithography. This is mainly because of the higher efficiency compared to the EBL or IBL technology, and has a wide workspace and low cost. LIL has the following advantages compared with other nanolithography technologies: (1) low cost (2) very high throughput; (3) no contamination on the surface; (4) capability to fabricate patterns large areas (up to hundreds of mm in diameter); (5) program controlled re-configurable patterns (with different periods, feature sizes and pattern shapes).

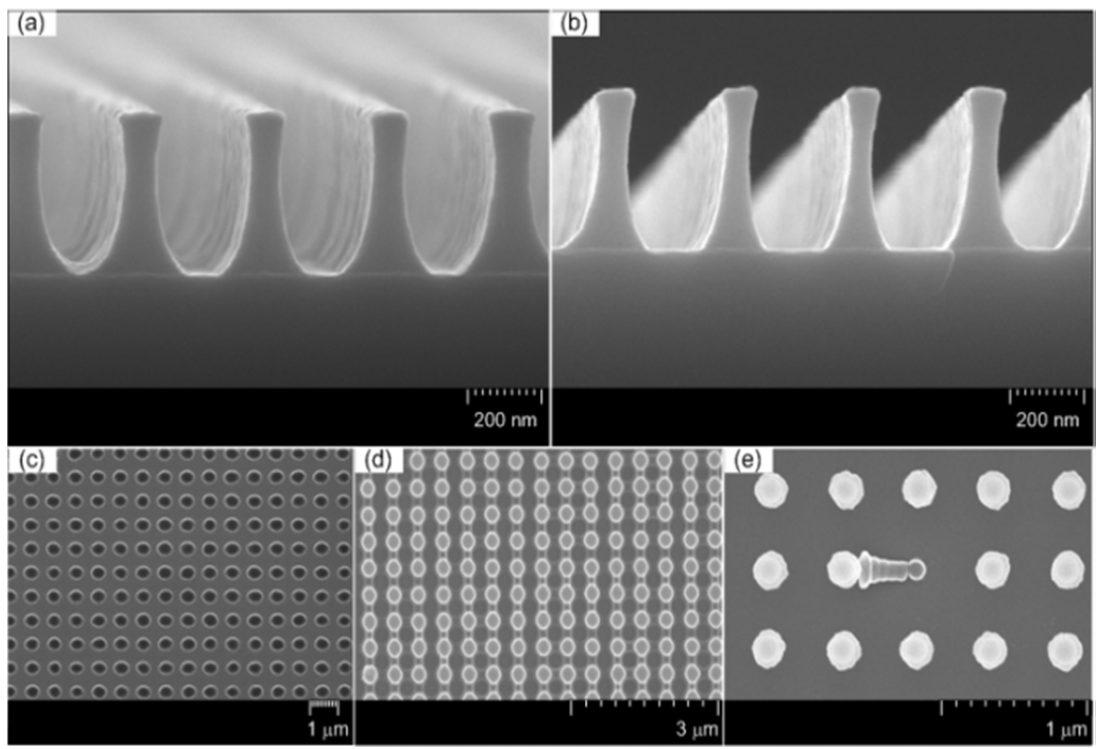

Fig. 1. SEM images of different exposure dosage. [31]

## 2. Laser Interference Lithography

2.1 Process parameters for LIL

Experimental characteristics could be defined by exploring the effects of several key parameters of LIL process, which could provide further understanding. These parameters include exposure dosage, beam power and the half angle between beams. In addition, optimization of the generated patterns is of great importance which could be achieved by the application of anti-reflective coating materials. To simplify the process, a single beam Lloyd's mirror interferometer was considered for the experiments [31] with 515 nm wavelength laser, frequency doubled by a beta-barium borate (BBO) crystal [32]. In order to define the required exposure dosage, case studies were performed with constant values of other parameters, such as LIL angle and beam power. It can be clearly seen that insufficient dosage resulted in holes (Fig. 1(c)) instead of columns, whereas excessive dosage generated inconsistent shape of patterns and could result in collapse of the structure (Fig. 1(d) and (e)), with the utilization of a positive photoresist.

Intensity distribution is another critical factor for fabricating uniform periodic patterns especially on large area, where it is a function of the wave amplitude of the partial beams, the wavelength of the light source, and the LIL angle. For a two-beam interference pattern, intensity $I(x)$ is given by [33]

$$I(x) = 2A\left\{\cos\left[\frac{4\pi x}{\lambda}\sin(\theta)\right]+1\right\} \quad (1)$$

where $A$ is the wave amplitude of the partial beams, $\lambda$ is the wavelength, and $\theta$ is the LIL angle. Utilizing the spatial filter to expand the beam could minimize the non-uniform intensity distribution, however, for lager area fabrication, higher expansion of the beam or longer distance to the sample may be required.

Followed by the definition of the required energy dosage, the effects of the half angle between two beams at the intersection were explored by the rotation of the stage and thus varying the angle θ. As mentioned previously, the relationship between the LIL angle (θ) and the pitch of the pattern (Λ) can be expressed as

$$\Lambda = \frac{\lambda}{2\sin\theta} \quad (2)$$

By observing the size of the pitch and the LIL angle, it is apparent that Fig. 2 agrees well with the above equation, where the smallest size of the pitch is half the wavelength of the laser.

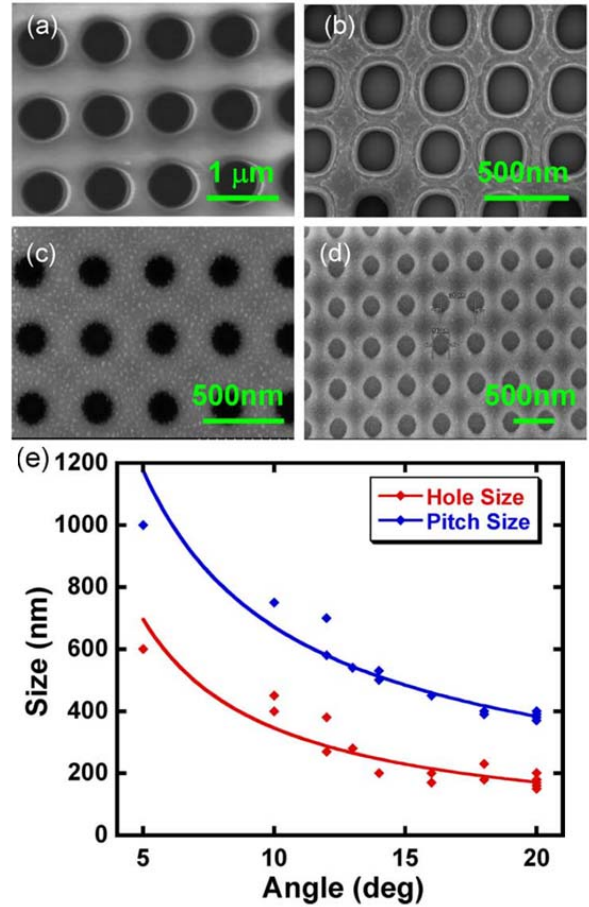

Fig. 2. (a-d) SEM images of transferred Si NM on glass substrate with light incident angles of 5, 10, 15, and 20 degrees, respectively. (e) Measured hole and pitch sizes as a function of the LIL angle. [66]

Moreover, the power of the beam could be increased to reduce the exposure time for quickening the LIL procedure. Although this seems to be apparent, since the total energy applied to the sample is identical, the difficulties in realizing large area uniformity could arise as the area of intensity distribution decreases.

## 2.2 Enhancement on pattern resolution

The gradual decrease in thickness of the nanoscale pillar, shown in Fig. 1(e), is due to the vertical standing waves caused by the reflection from the bare silicon substrate [34]. To prevent generation of undesired patterns, anti-reflective coating is widely used where it controls the reflection of the laser light source at the surface of the substrate [35-38].

Fig. 3 shows the comparison between samples with and without the application of bottom anti-reflective coating materials (BARC) for one and two-dimensional patterns. The exposure dosage was kept constant, and identical experimental procedures were performed for each case. For the samples with the photoresist only, shown in left sides of Fig. 3, inconsistent shapes could be observed. As the vertical standing waves were caused from the reflection from the Si substrate, the width of the pillar was continuously reduced as it reaches closer to the substrate and irregular wall surface could also be observed. By applying the BARC materials, improvements on the patterns could be clearly observed in right side of Fig. 3.

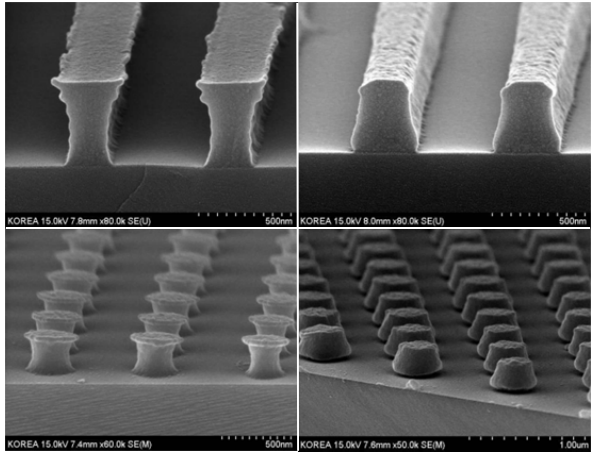

Fig. 3. SEM images of one and two dimensional patterns without BARC (left); and with BARC(right). [35]

With the introduction of the BARC materials, the patterns show more uniformity and stability, improved conditions for post processing, inconsistent shapes and irregular wall surfaces reduced, preventing the possibilities of the collapse of the structures. Additionally, pattern footing and T-top phenomena from the development process were significantly reduced.

Through application of the BARC materials, it is possible to generate smaller uniform patterns, enhancing the pattern resolution and size limitations of LIL.

Another important issue concerning LIL is clearly the pattern size limitation. Although short wavelength light source could be used and therefore minimize the generated feature sizes, further investigation on the reduction of pattern to pitch size ratio, less than half of the pitch length, could provide further understanding of the process [35].

In order to reduce the pattern size by increasing the exposure energy, the LIL angle was set constant and the BARC materials were applied for uniform and stable pattern fabrication. Ar-Ion laser (257 nm wavelength) was utilized as the light source and the pitch size was set to be around 770 nm, where the energy doses varied from 9 to 18 mJ/cm$^2$. Fig. 4 shows the resulting top and cross-sectional views of the fabricated line patterns.

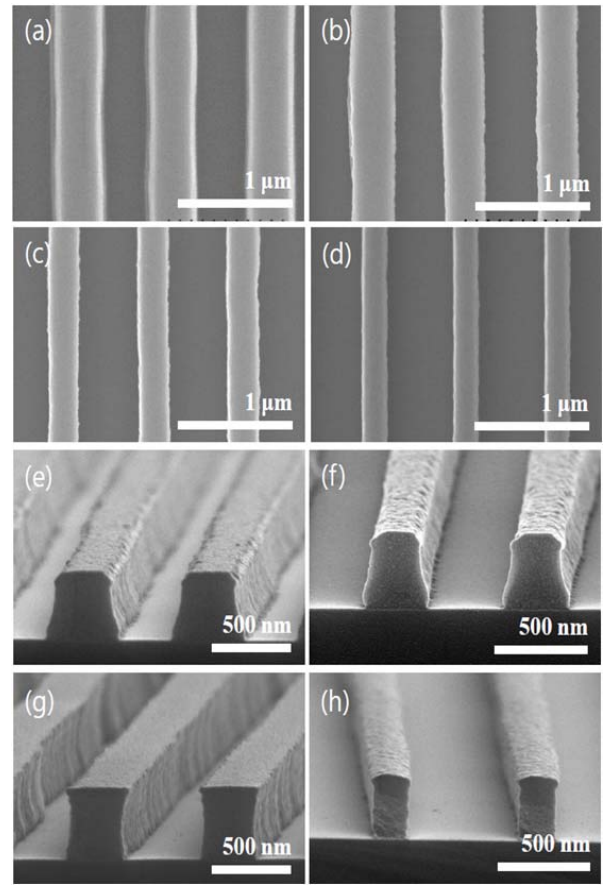

Fig. 4. SEM images of patterns at different exposure energy doses (a)(e) 9, (b)(f) 12, (c)(g) 13.5, and (d)(h) 18 mJ/cm2. [35]

Table 1. Pitch and pattern sizes at different energy doses applied. [35]

| Energy dose (mJ/cm2) | Pitch Size (nm) | Pattern Size (nm) | Ratio (Pitch: Pattern) |
|---|---|---|---|
| 9 | ~770 | 390 | 1.97 : 1 |
| 12 | | 340 | 2.27 : 1 |
| 13.5 | | 290 | 2.66 : 1 |
| 18 | | 190 | 4.05 : 1 |

The measured pattern sizes are itemized in Table 1. It could be clearly observed that increasing the exposure energy results in reduction of pattern size, reaching less than the quarter of the pitch size in this experiment. The inverse linear relationship between energy dose and the pattern size was also confirmed while maintaining the stability and uniformity of the patterns. These results highlight the possibility of pattern reduction which could be applied for various short wavelength light sources to generate minimal sized features and offer different geometrical designs for post procedures such as reactive ion etching and photoresist lift-off.

As mentioned above, LIL is a technique which involves a maskless process suitable for large area patterning. Although this capability is one of the main advantages for most LIL applications, optional selective patterning is often required for many applications. To account for the non-selective patterning limitation of LIL, design and development of combined LIL and photolithography have been reported [39].

Combined LIL and photolithography patterning technique consists of reactive ion etching in between the LIL and the photolithography process to fabricate a hybrid mask mold containing micro and nanopatterns. Introduction of a sacrificial layer is required for nanoscale patterns and of a photomask for microscale patterns.

The possibility of pattern resolution enhancement utilizing BARC, variation of exposure energy, and selective patterning could lead to various applications, especially for device production.

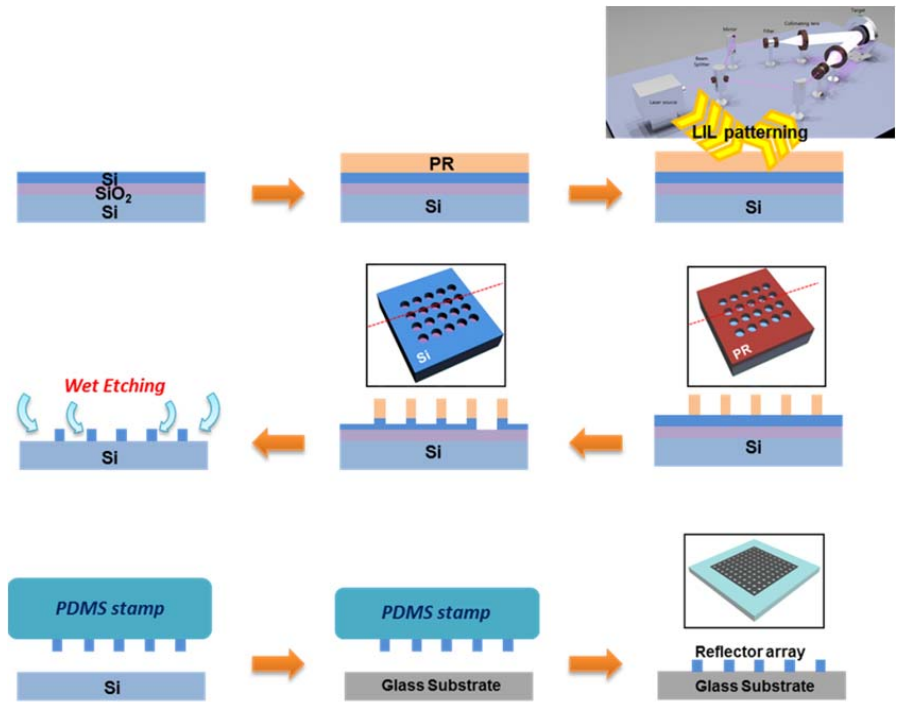

Fig. 5. Schematical illustration of fabrication process for Si-MR on the glass substrate using transfer printing technique with assistant of elastomeric stamp.

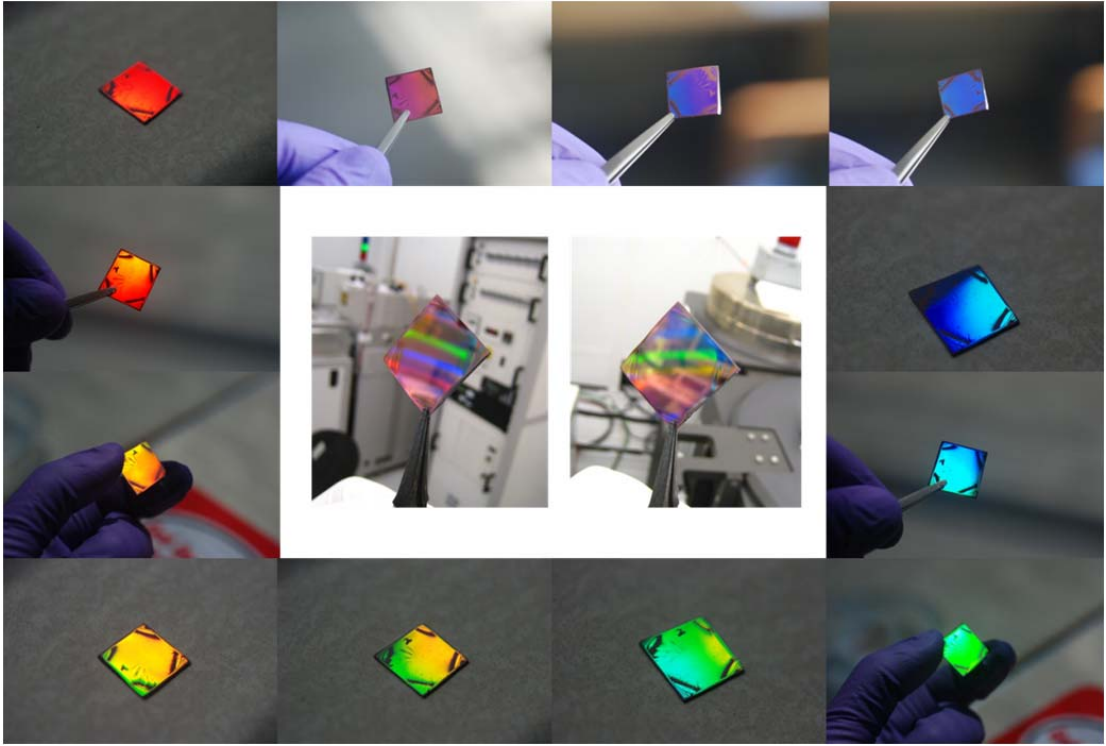

Fig. 6 Sample images after LIL pattern followed by RIE etching on SOI wafer.

## 3. Application of optical devices by LIL

3.1 Large area photonic applications

Two-dimensional (2D) photonic crystals slabs (PCSs) are one of the most promising artificial platforms for optical applications such as microcavities, lasers, photodetectors, solar cells, sensors, reconfigurable photonics, and etc [40, 41]. Many of conventional nanopatterning techniques involve a slow process and are not cost effective. LIL, on the other hands, enables to generate periodic patterns faster on the larger area than conventional methods, and thus is suitable to create 2D PCS structure. With this point of view, fabrication of PCSs with LIL suggests the easiest method to achieve large-size PCSs without losing performance aspect. Most of 2D PCSs were made by further deposition or etching to the substrate using photoresist patterns defined by LIL [42, 43]. Therefore, 2D PCSs by LIL should make on either single crystalline substrate and form a homogeneous layer structure or amorphous/polycrystalline layer structure by deposition followed by lift-off. For the former case, it is difficult to take advantage from refractive index difference between PCS layer and substrate and for the latter case, strong absorption loss, due to the defects in amorphous/polycrystalline material, often degrades overall device performance. As a result, this limitation hinders toward more delicate and complex photonics system. The recent advances of releasable and transfer printable Si nanomembranes (NMs)[44-53], still inheriting the single crystal quality of bulk Si, have enabled Si to be applicable to optical devices built on the foreign substrate by printing method [54-59]. While there're many photonic crystals made with polymers, amorphous materials, and metals [60-65], Si NMs have very high index but the absorption loss is almost negligible, therefore, with Si NMs, a unique and high performance 2D PCS can be realized. In this review, we introduce 2D Si NM PC mirrors built on the glass substrate by transfer printing method as an example [66].

Fig. 5 shows the typical of fabrication process for 2D PCSs built on a glass substrate with Si NM. Fabrication begins with patterning nanostructures on the SOI wafer (SOITEC$^{TM}$). Top Si with the photoresist as etching mask was etched by deep reactive ion etching (DRIE), followed by the selective undercut etching of the buried oxide layer. The released patterned top Si layer is, then, referred to as Si membrane reflector (Si-MR). This Si-MRs were picked up and printed onto the glass substrate by employing a polydimethylsiloxane (PDMS) stamp printing technique without using any adhesive layer between the Si-MR and the glass substrate [67]. After transfer, the sample was annealed at 220 °C for 180 sec using RTA under nitrogen atmosphere in order to improve the bonding strength between the transferred Si-MR and the glass substrate.

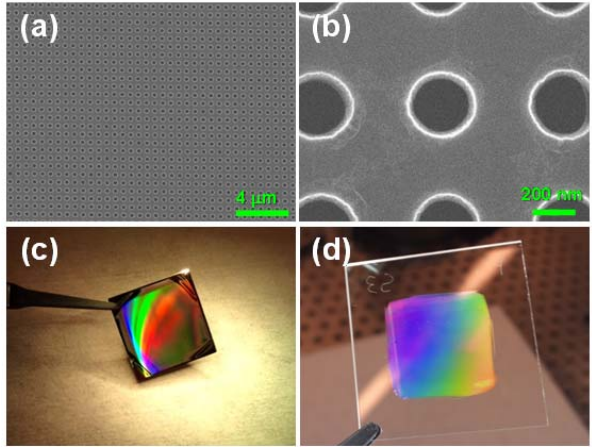
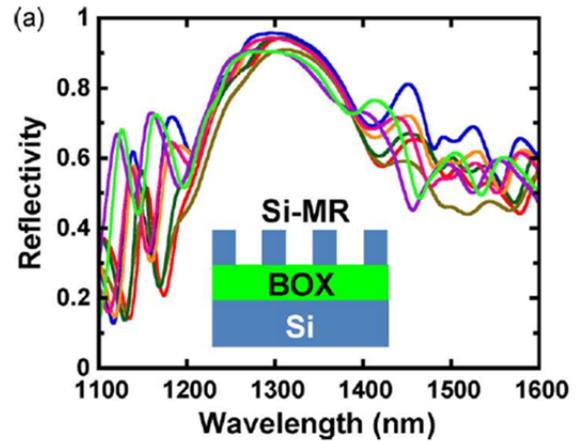

Fig. 7. (a) SEM image of the hole array pattern formed on a SOI substrate. (b) Zoomed-in image of (a). (c) Optical image of an LIL-patterned sample after dry oxide (BOX) layer by immersing the sample in concentrated hydrofluoric acid (HF, 49%). etching of a top Si layer on SOI. (d) Optical image of a transferred Si-MR on a glass substrate. [66]

Fig. 7(a) and (b) show the SEM images taken on the Si-MR after printing on the glass substrate. The dimensions of nanopatterns are 240 nm hole radius with 500 nm pitch distance. Shown in Fig. 7(c) and (d) are sample images of a 2 cm × 2 cm sized Si-MR, before and after printing onto the glass substrate. With the PDMS-assisted printing process, large-area and high-quality Si NMs have been successfully transferred without any visible fractures and finally Si-MR was fabricated on the foreign substrate (a glass substrate).

The reflectivity of the MR was measured at normal incidence with a beam size of around 100 μm. The spectra are shown in Fig. 8(a) and (b), for the reflectors before print (on SOI wafer) and after print (on the glass), respectively. The reflections of these Si-MRs at different locations are also tested, and the results are plotted together using different colors.

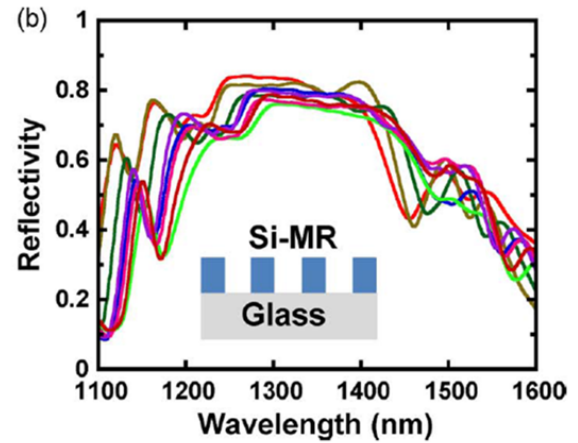

Fig. 8. Reflection spectra measured at different locations of the large-area Si-MR: (a) measured spectra of Si-MR on SOI wafer before undercutting of the BOX layer and (b) measured spectra of Si-MR after onto a glass substrate.

As shown in Fig. 8(a), the Si-MR peak reflectivity on SOI is as high as 95% at a near infrared wavelength range (~1300 nm) with good uniformity at different locations. The bandwidth is about 100 nm from 1250 to 1350 nm. Theoretically, 100% reflection can be achieved, which can be realized with further fabrication optimizations.

On the other hand, some have used LIL by forming a periodically structured array to decrease light reflection from the surface. [68, 75] Previously, patterning by e-beam lithography or spreading nanospheres, followed by dry-etching are the most popular methods to generate reasonably high anti-reflective coating layer [69-75].

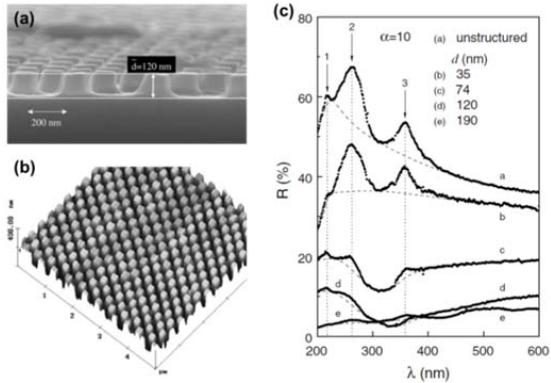

Fig. 9. (a) A cross sectional view of SEM of anti-reflective coating on Si, (b) A tilted atomic force microscope (AFM) image of anti-reflective coating on Si, (c) the reflectivity from anti-reflective coating with various depth. [75]

But LIL can be easily applied to fabricate large area anti-reflective coating. Theoretically, reflection reduces as the layer thickness over the target wavelength increases [69], thus anti-reflection at certain wavelength can be easily realized by the relationship shown in Fig. 2. Hadobas et al. demonstrates anti-reflective coating layer on Si substrate and about 5% of reflection in visible range was achieved as shown in Fig. 9 [75].

3.2 Plasmonic color filter

The "plasmonics" has been gathering much interest of many scientists due to its interdisciplinary characteristics. The plasmonic structures accompany an optical resonance phenomenon, which is called surface plasmon resonance (SPR), the free electrons oscillate at the surface of metal. Researchers have been amassing the ways to handle the metal in nano scale structures with various nanofabrication methods, such as e-beam lithography, ion milling, self-assembly, interference lithography, and so on. That has led advances in research approaches with the SPR by suggesting various applications. [76-81]

The color filters widely used for the industrial devices such as organic light emitting diodes (OLEDs), Liquid Crystal Displays (LCDs), and CMOS image sensors, are composed of organic dyes (or pigments). The color-reproduction performances of these dye-based filters are highly dependent on the material intrinsic characteristics; color resist in the red filter, for example, absorbs all wavelength regions except the light in red range. The filtering performance originated from color sensitivity of dyes is degraded by heat and ultraviolet radiation due to low chemical stability of the organic materials [82, 83].

On the other hand, the PCFs are combining an optically thin metal layer, and its transmittance can be tuned by the geometrical and material conditions: the periodicity, size and shape of holes, the thickness of metal, and the optical constants of materials. This simple and thin structure is advantageous to be assembled into other devices regardless of degradation by heat and light. Many reported results in earlier stage have demonstrated PCFs as indicating easy tunability of transmittance [84-87]. More recently, experimental analysis about spatial cross talk and the effect of defect was done in detail, and showed more possibility to industrial applications [79]. However, the fabrication method used up to now, such as nanoimprinting [84, 88-90], electron beam lithography [76-81] or focused ion beam [91-95], restricts mass-production of PCFs, encountering problems of low speed, small patterning area, and high cost equipment. Here we suggest a fabrication flow including laser interference lithography (LIL) step. On the contrary to above-mentioned technologies, LIL yields perfect ordering patterns, which must be spatially coherent over large area, with simple maskless equipment. Although LIL has the limitation that it can only fabricate simple periodic patterns, it is an attractive additional solution of conventional methods for applications in which periodic patterns are desirable, including x-ray transmission gratings, photonic crystals and sub-micrometric sieves. In this regard, fabrication of plasmonic color filters with LIL enables to suggest the easiest method to achieve large size PCFs without losing performance aspect. In this work, we demonstrate PCFs with primary colors (the red, green, and blue) on the area of 2.5 cm × 2.5 cm. The transmittance of the filters was optimized by analyzing SP modes at the interfaces between Aluminum (Al) and Lithium Florid (LiF).

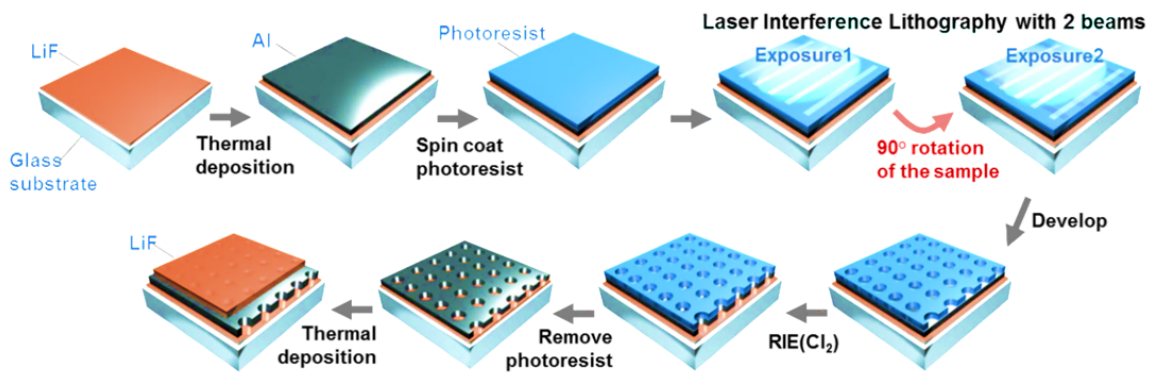

Fig. 10. Schematic diagram of the proposed fabrication process. In a Lloyd's mirror interferometer system, two laser beams form a 1D interference pattern on the specimen. To create array patterns with a square symmetry, two exposures by a rotation of 90° in between are applied. [103]

Fig. 10 shows the suggested fabrication procedure. A LiF layer (50 nm) and an Al layer (150 nm for the red and green; and 100 nm for the blue) were thermally evaporated in sequence on the glass substrate. In the following, laser inference lithography was performed with a Lloyd's mirror interferometer system [19, 35, 66]. To make the interference patterns, two beams were used: one travels directly to the specimen and the other was reflected onto the specimen by the mirror. These two beams formed a 1D interference pattern on the specimen [96-98]. The optimized dimensions for each structure are specified in Table 2. The last column indicates the average hole size, which was patterned on the photoresist. The size of the hole can be tuned by adjusting the exposure dosage, the beam power and the time for developing. In order to optimize the transmission characteristics, structures consisting of a unit cell were simulated by three-dimensional (3D) finite-difference time-domain method (FDTD, Lumerical solutions, Canada). The optical constants of the glass substrate and the Al were based on the Palik data; the optical constant of LiF was determined from the experimental data. A plane wave source polarized in the x direction was launched at normal incidence.

As shown in Figure 11b, the $SP_{b(1,1)}$ and $SP_{u(1,1)}$ modes occurred simultaneously at $\lambda_3$, similarly to the $SP_{b(1,0)}$ and $SP_{u(1,0)}$ modes at $\lambda_1$ (Figure 11d), resulting in a similar amount of strong field distribution at the top and bottom of the hole. The $SP_u$ mode at $\lambda_2$ appeared to be much weaker than it was at $\lambda_3$ and $\lambda_1$. We supposed that the top of the structure became planar when the upper LiF was deposited in a layer more than 150 nm thick. Figure 11e-h illustrates the response of the structure with the upper LiF layer of 200 nm, indicating that the matched $SP_b$ and $SP_u$ modes were maintained after planarization.

Table 2. The dimensions of the optimized structure for each color. The average values of the diameter of the holes in the fabricated samples were rounded off to the tens place. [103]

|  | Period (P) [nm] | Thickness of Al [nm] | Diameter of holes [nm] | Diameter of holes (Experimental) [nm] |
| --- | --- | --- | --- | --- |
| Red | 390 | 150 | 200~230 | ~210 |
| Green | 320 | 150 | 180~200 | ~180 |
| Blue | 230 | 100 | 120~160 | ~120 |

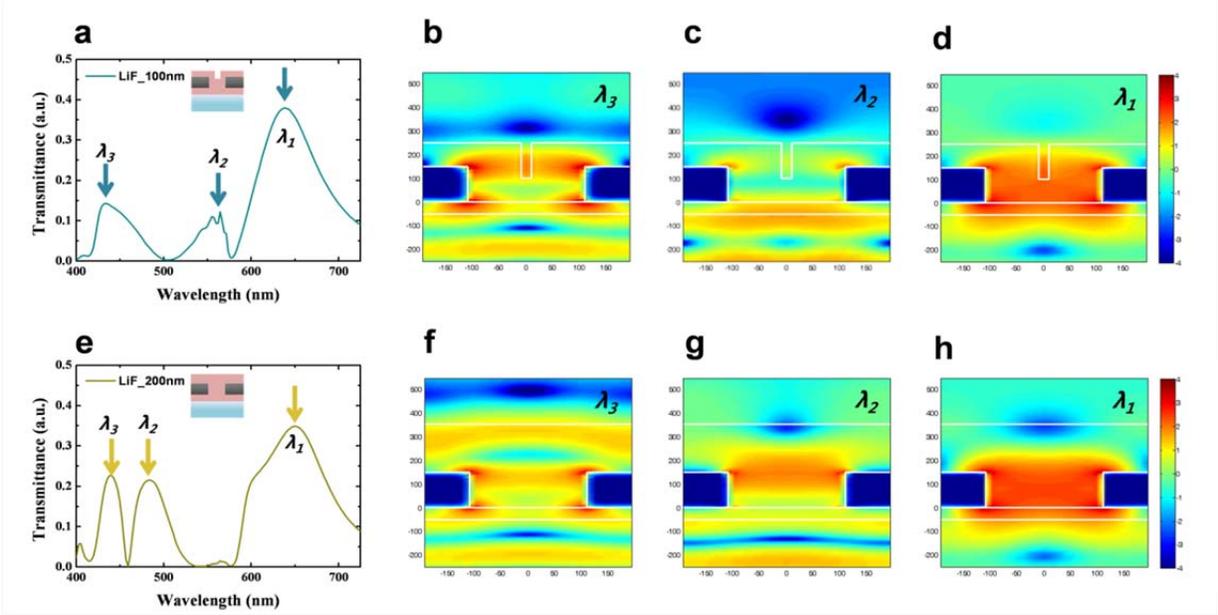

Fig. 11. Transmittance of the PCF of which the cylindrical surface as well as the top surface are covered with LiF of (a) 100 nm and (e) 200 nm, respectively. (b), (c), and (d) illustrate the electric field intensity, $\log(|E|^2)$, at the wavelength of $\lambda_3$, $\lambda_2$, and $\lambda_1$ in (a). (f), (g), and (h) illustrate the electric field intensity, $\log(|E|^2)$, at the wavelength of $\lambda_3$, $\lambda_2$, and $\lambda_1$ in (e). [103]

The thickness of the LiF at the bottom of the metal-dielectric interface had a relatively small effect on the $SP_b$ modes. Because the thickness of the LiF was scaled to the penetration depth of the surface plasmon ($\delta_d$), the thickness affected the effective permittivity ($\varepsilon_d'$), which ranged from the permittivity of: 1) the air to LiF in the case of the upper LiF, 2) LiF to $SiO_2$ for the bottom. The difference in permittivity between LiF and $SiO_2$ was less than that between LiF and the air: 0.23 for the former one; 0.93 for the latter, at a 400 nm wavelength. Therefore, the LiF between the substrate and Al had less influence on the transmittance, as shown in Figure 12.

Fig. 13(a) provides the photograph of the PCFs. The color gamut of each filter was mapped on the CIE 1931 xy chromaticity diagram (Fig. 13(b)). As with the transmittance results, the red and blue filters showed worse color purity compared to that of the green filter. Since Z, which is quasi-equal to the blue stimulation, is the biggest among the tristimulus values, the blue-noise light can have the worst effect on the color purity. In the hexagonal arrays, the transmittance peaks originating from the 1st and 2nd resonance modes are split into further position in wavelength region than are the square arrays [100-102]. This would be advantageous for producing pure red color. Fortunately, three laser beams make a 2D-interference pattern with a hexagonal array, so there is chance, experimentally, to improve the color purity.

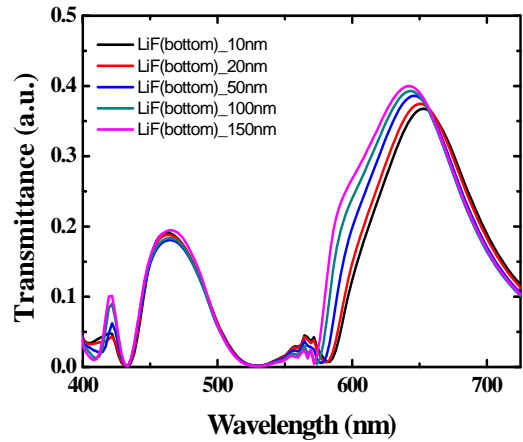

Fig. 12. Transmittance of the PCFs according to the thickness of the LiF layer encountering the bottom of the Al film

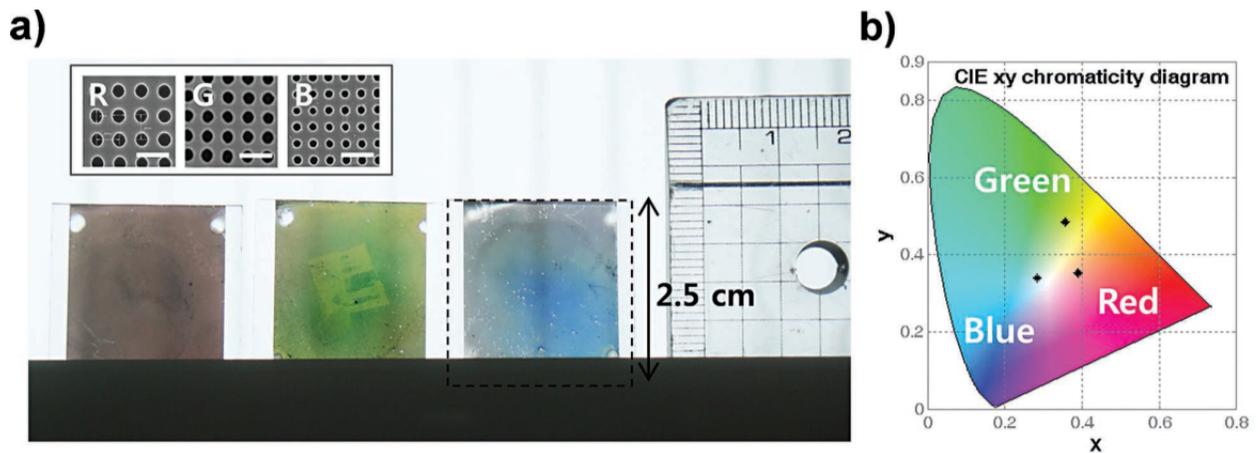

Fig. 13. Chromaticity characteristics of the PCFs. (a) Photograph of PCFs: the red, green, and blue from the left. The vertical blind in the background shows through the filters. Dashed line represents the boundary of the sample. Insets are the SEM images of the photoresist of each filter, with a white scale bar of 500 nm. (b) CIE chromaticity diagram of the fabricated PCFs. Each color corresponds to the point of (x, y): (0.3896, 0.3523) for the red, (0.3559, 0.4841) for the green, and (0.2836, 0.3384) for the blue. [103]

## 4. Conclusion

Among various nanoscale patterning techniques, LIL could be advantageous for its maskless, quick, and simple process characteristics. By exploring the experimental aspects of LIL, the advantages and limitations are highlighted for generating nanoscale patterns. Definitions and the effects of the critical process parameters, such as exposure dosage, intensity distribution, LIL angle, and beam power, have been discussed followed by pattern resolution improvement using anti-reflective coating materials. Minimizing the pattern size through variation of exposure energy and introduction of combined LIL and photolithography technique to account for non-selective patterning limitation of LIL has been discussed.

Optical/photonic applications by LIL indicate that simple, fast, and large area processable device can be realized. We have reviewed various periodic structures such as reflectors and anti-reflectors which reflection is over 95% at a near infrared wavelength range for reflectors and which reflection is less than 5% for anti-reflective coating layers, respectively. This implies LIL can produce key components in optical system with relatively cheap price. We have also reviewed large area PCFs which are effectively colored as R, G, B. The results show great potential of PCFs to be applicable to the display applications. The LIL process provides nano-scaled array patterns through whole area of 2.5 cm × 2.5 cm and obtained 20.0%, 19.9%, and 22.9% for the maximum transmittance of the red, green, and blue, respectively.

We expect that these optical/photonic applications will provide the opportunity to lead more advanced applications to mass production.

## Acknowledgment

This work was supported in part by the U.S. AFOSR under Grants FA9550-091-0482, FA9550-08-1-0337, and FA9550-11-C-0026 and the IT R&D program of MKE/KEIT(Grant No. 10041416, The core technology development of light and space adaptable new mode display for energy saving on 7inch and 2W) and by the Korea Institute of Science and Technology (2E23892). Zhenqiang Ma (mazq@engr.wisc.edu), Jinnil Choi (jlchoi@hanbat.ac.kr) and Byeong-Kwon Ju (bkju@korea.ac.kr) are co-corresponding authors.

**Graphical Abstract**

A systematic review, covering fabrication of nanoscale patterns by laser interference lithography (LIL) and their applications for optical devices are provided. LIL is a patterning method with simple, quick process over a large area without using a mask which is the ideal method to create optical and photonic nano structures. Demonstration of optical and photonic devices by LIL is reviewed such as directed nano photonics and surface plasmon resonance (SPR) or large area membrane reflectors and anti-reflectors. Perspective on future directions for LIL and emerging applications in other fields are presented.

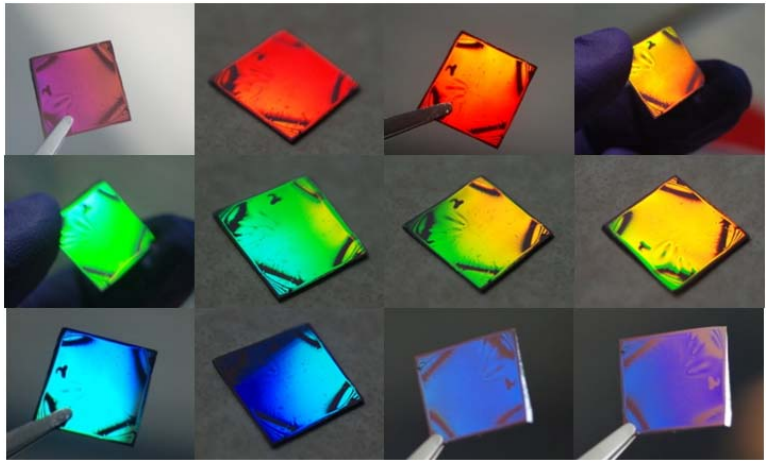